\documentclass{PoS}
\usepackage{url}
\usepackage{color}
\usepackage{subfigure} 
\usepackage{multicol} 
\usepackage{nonfloat}
\usepackage{caption}
\usepackage{lineno}

\newcommand\myfigure[1]{%
\medskip\noindent\begin{minipage}{\columnwidth}
\centering%
#1%
\end{minipage}\medskip}

\title{Development of the LAGO Project in Chiapas-Mexico}

\ShortTitle{}

\author{\speaker{Karen Salom\'e Caballero Mora} $^1$, Hugo de Le\'on Hidalgo $^1$, Eduardo Moreno Barbosa$^{2}$, C\'esar \'Alvarez Ochoa$^1$, Roberto Arceo Reyes$^1$, Filiberto Hueyotl Zahuantitla$^1$, Sarah Kaufmann$^3$, Luis Rodolfo P\'erez S\'anchez$^{1,3}$, El\'i Santos Rodr\'iguez$^{3}$, Omar Tibolla$^{3}$ and Arnulfo Zepeda Dom\'inguez$^{3,4}$ for the LAGO Collaboration $^5$\\
      \llap {$^1$}  Facultad de Ciencias en F\'isica y Matem\'aticas, Universidad Aut\'onoma de Chiapas (FCFM-UNACH)\\
Carretera Emiliano Zapata Km 8, Rancho San Francisco, Ciudad Universitaria, Ter\'an, Tuxtla Guti\'errez, Chiapas, C.P. 29050, Mexico\\
    \llap {$^2$} Meritorious Autonomous University of Puebla (BUAP) \\
4 sur 104 Centro Hist\'orico, Puebla, C.P. 72000, Mexico \\
     \llap {$^3$}  Mesoamerican Centre for Theorethical Physics (MCTP-UNACH)\\
Carretera Emiliano Zapata Km 8.4, Rancho San Francisco, Ciudad Universitaria, Ter\'an, Tuxtla Guti\'errez, Chiapas, C.P. 29050, Mexico\\
    \llap {$^4$}  Center for Research and Advanced Studies of the National Polytechnic Institute (CINVESTAV)\\
Av. Instituto Polit\'ecnico Nacional 2508, Col. San Pedro Zacatenco, Delegaci\'on Gustavo A. Madero, México D.F. C.P. 07360. \\
\llap {$^5$} The Latin American Giant Observatory (LAGO), \textcolor {blue}{\url {http://lagoproject.org/}} \\
See all members at \textcolor {blue}{\url {http://lagoproject.org/collab.html}}\\
        E-mail: \email{karen.scm@gmail.com}}

\abstract{The Latin American Giant Observatory (LAGO) is an extended astroparticle observatory with the goal of studying Gamma Ray Bursts (among other extreme universe phenomena), space weather and atmospheric radiation at ground level. It consists of a network of several Water Cherenkov Detectors (WCD) located at different sites and different latitudes along the American Continent (from Mexico up to the Antarctic region). Another interest of LAGO is to encourage and support the development of experimental basic research in Latin America, mainly with low cost equipment. In the case of Chiapas, Mexico, the experimental astroparticle physics activity was limited, up to now, to data analysis from other detectors located far away from the region. Thanks to the collaboration within LAGO, the deployment of one WCD is ongoing at the Universidad Aut\'onoma de Chiapas (UNACH). This will allow, for the first time in the region, to train students and researchers in the deployment processes. Till now the setup of the signal-processing electronics has been performed and the characterization of the photomultiplier tube is currently being done. The main, short-term goal is to install one WCD on top of the Tacan\'a volcano in Chiapas in a short term. The status of the work is presented.}

\FullConference{35th International Cosmic Ray Conference --- ICRC2017\\
		10--20 July, 2017\\
		Bexco, Busan, Korea}

\begin{document}

\section{Introduction}

The Latin American Giant Observatory (LAGO) is a global scale astroparticle observatory consisting of an integrated detection network. LAGO was originated in 2005, proposed by a group of astroparticle physicists who were members of the Pierre Auger Cosmic Ray Observatory \cite {Auger}. The project aims to install small water Cherenkov detectors at  high altitude ( > 4500 m a.s.l.) in order to detect the highest energy components of Gamma Ray Bursts (GRB) \cite{Lago}, \cite{Conde}, \cite{arxiv}. The LAGO Collaboration is integrated by about 100 researchers and students from 28 institutions from 11 countries in Latin America. UNACH is part of the collaboration through the FCFM and the MCTP. Furthermore, the Chiapas group has the support of researchers from CINVESTAV and BUAP. The main objectives of the project are: 1.- Scientific objectives: to study high energy astroparticles, Meteorology and Space Climatology, and Atmospheric radiation and its applications 2.- Academic objectives: To train Latin American students in high energy physics and astroparticle physics, and to form an open and collaborative network of high energy physics researchers. Figure \ref{fig:1} shows a map of places where there is or there will be a LAGO site.

\begin{multicols}{2}

\myfigure{\includegraphics[scale=.47]{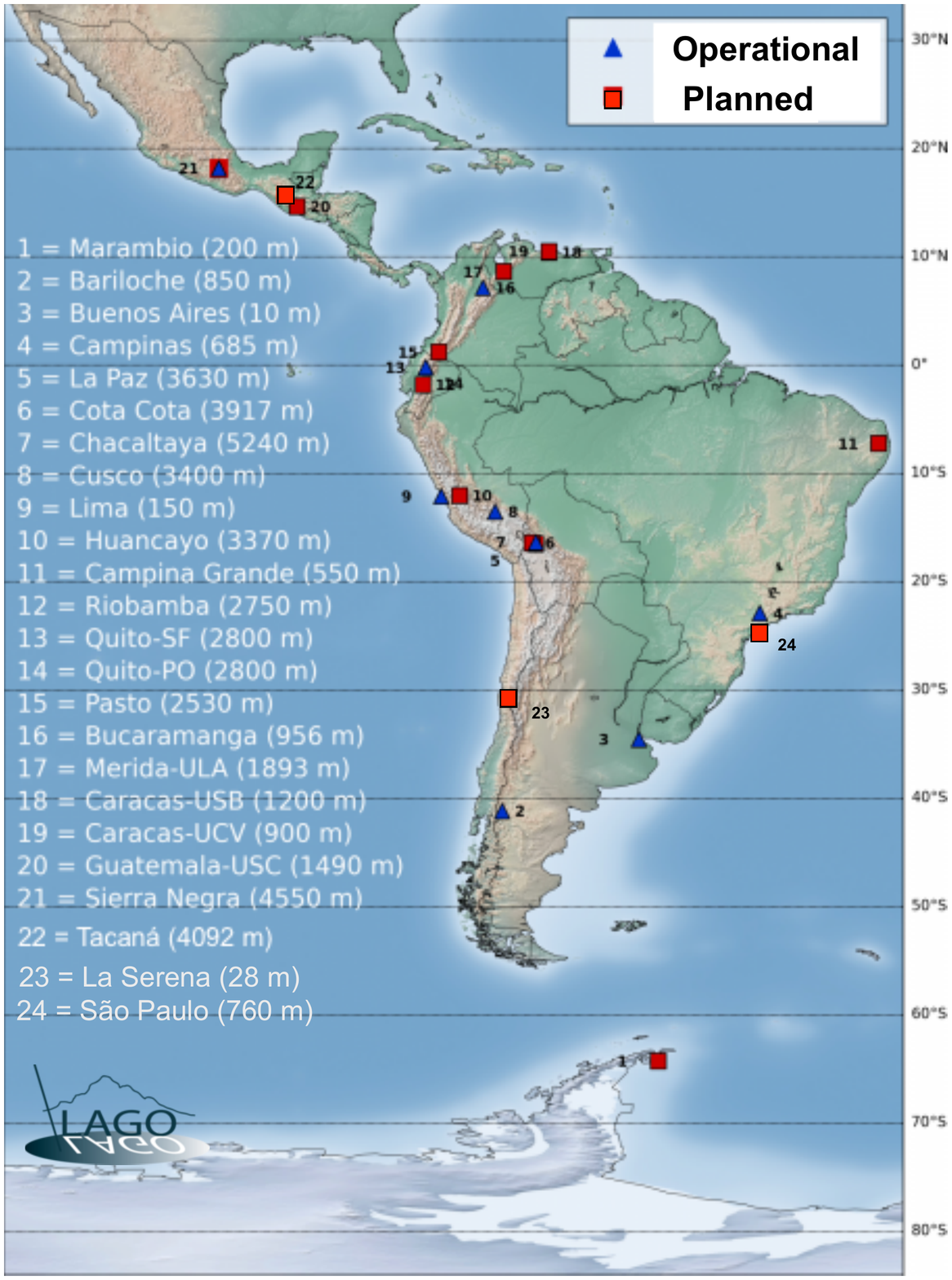}%
\figcaption{\emph{LAGO sites, modified from \cite{Lago}.}}}
\label{fig:1}
\hspace{2.0pc}
\myfigure{\includegraphics[scale=1.31]{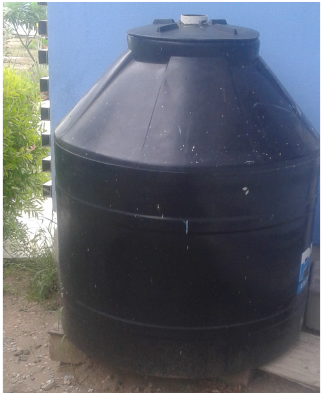}%
\figcaption{\emph{Test tank at Chiapas site.}}}
\label{fig:2}
\hspace{2.0pc}
\myfigure{\includegraphics[scale=1.33]{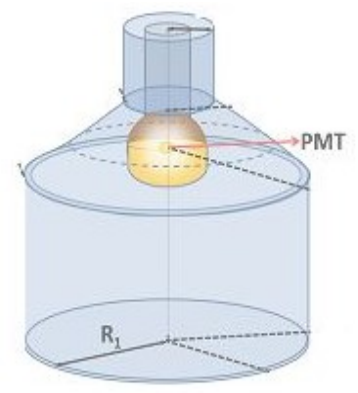}%
\captionsetup{width=.8\linewidth}
\figcaption{\emph{Schema of the inside of the tank, modified from \cite{Perez}.}}}
\label{fig:3}

\end{multicols}

\section{Methods}
The LAGO detection network consists of particle detectors, set up individually or forming small networks. The network covers a very wide latitudinal distribution, with detectors installed from Mexico to Patagonia, and will soon cover the Antarctic region. Located mostly near the Andes Mountains, the detectors are placed at a wide range of altitudes: from sea level, with detectors in Lima (Peru) and Buenos Aires (Argentina), to more than 5000 m above the level of the sea in the Nevado de Chacaltaya (Bolivia). This distribution allows to cover an extensive range of geomagnetic rigidities and atmospheric absorption and reaction levels \cite{Lago}. 

In the case of Chiapas, the initial site on which the experiment will be deployed and tested is the FCFM campus. A prototype tank is now available for testing (see \figurename{ 2}). For the acquisition of the signals photomultiplier tubes (PMTs) are used, one per tank, located on the top of it (see \figurename{ 3}). The corresponding signals are acquired through a data acquisition card. The card used for the first setup is type CAEN which follows the VME64 computer standard. At the moment, there are two types of PMTs, Photonis and Hamamatsu, available. 
The basic idea of the data acquisition system is to constantly monitor the signal delivered by the PMT so that if at a given  moment the signal exceeds a certain threshold, which indicates the detection of a high energy particle, it is recorded in a file for later count of the pulses. Saved files should include the time at which the detection happened and the duration times, which allow comparison with detections at other locations. In addition, it aims to make the information accessible remotely through the Internet.

At  MCTP there is another scintillator detector called Escaramujo \cite{Escaramujo} (see \figurename{ 4}) to measure  the passage of atmospheric muons. It uses plastic scintillators  and silicon photomultipliers SiPM (see \figurename{ 5}), the output is read with a time-to-digital converter board (QuarkNet \cite{quarknet}) . 

\begin{multicols}{2}

\myfigure{\includegraphics[scale=.65]{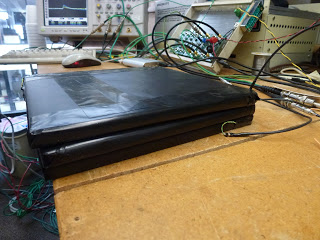}%
\figcaption{\emph{Escaramujo detector:three  25x25x1 $cm^3$ EJ-200 scintillator plates, wrapped with EMI/Static paper and Tyvek \cite{Escaramujo}.}}}
\label{fig:4}
\hspace{2.6pc}
\centering
\myfigure{\includegraphics[scale=1.85]{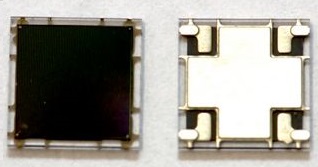}%
\figcaption{\emph{6x6 $mm$ MicroFC-60035-SMT SensL silicon photomultiplier, able to detect 1 single photon \cite{Escaramujo}.}}}
\label{fig5}

\end{multicols}
That instrument will be necessary to calibrate the PMTs that will be used in the tank. The final plan is to locate a tank at the Tacan\'a volcano (4092 m a.s.l.) on the border with Guatemala, a country which is also part of the LAGO collaboration. Once the detector is installed, the Chiapas group will contribute to the LAGO network to share, verify and complement measurements of events caused by emissions of gamma rays from the universe. Such measurements, performed with an aperture as large as that of LAGO, may relate the air showers generated by the astroparticles to highly energetic cosmic objects that can produce them, like GRB. The measurement range of LAGO is of the order of energies of 100 GeV. Detection of such particles is possible using the Single Particle Technique (SPT) \cite{presentacion}. The main idea in this detection technique is to know the particle count in a certain period of time, with this, it is possible to study the variations at another time. Other phenomena that can cause variations in measurements of particle count are variations in atmospheric pressure, electrical storms, solar activity, earthquakes and variations in the Earth's magnetic field, which should also be considered. The technique is applied using Cherenkov light detectors in water, as already described, and correlating their measurements with satellite records, observing fluctuations in the temporal measurements of the background radiation of each detector. These fluctuations will be caused by the arrival of secondary particles, mainly photons \cite{Nava}.

\section{Results}
1.- An electronic acquisition card has been set up and communication with the PMT has been established allowing the observation of signals. 2. Set up of the tank has been done: Tyvek has been deployed internally in the tank and the external wall has been covered with plastic and alluminium to isolate the PMT, avoiding interference from outside (see \figurename{ 6} and \figurename{ 7}).

\begin{multicols}{2}

\myfigure{\includegraphics[scale=.05]{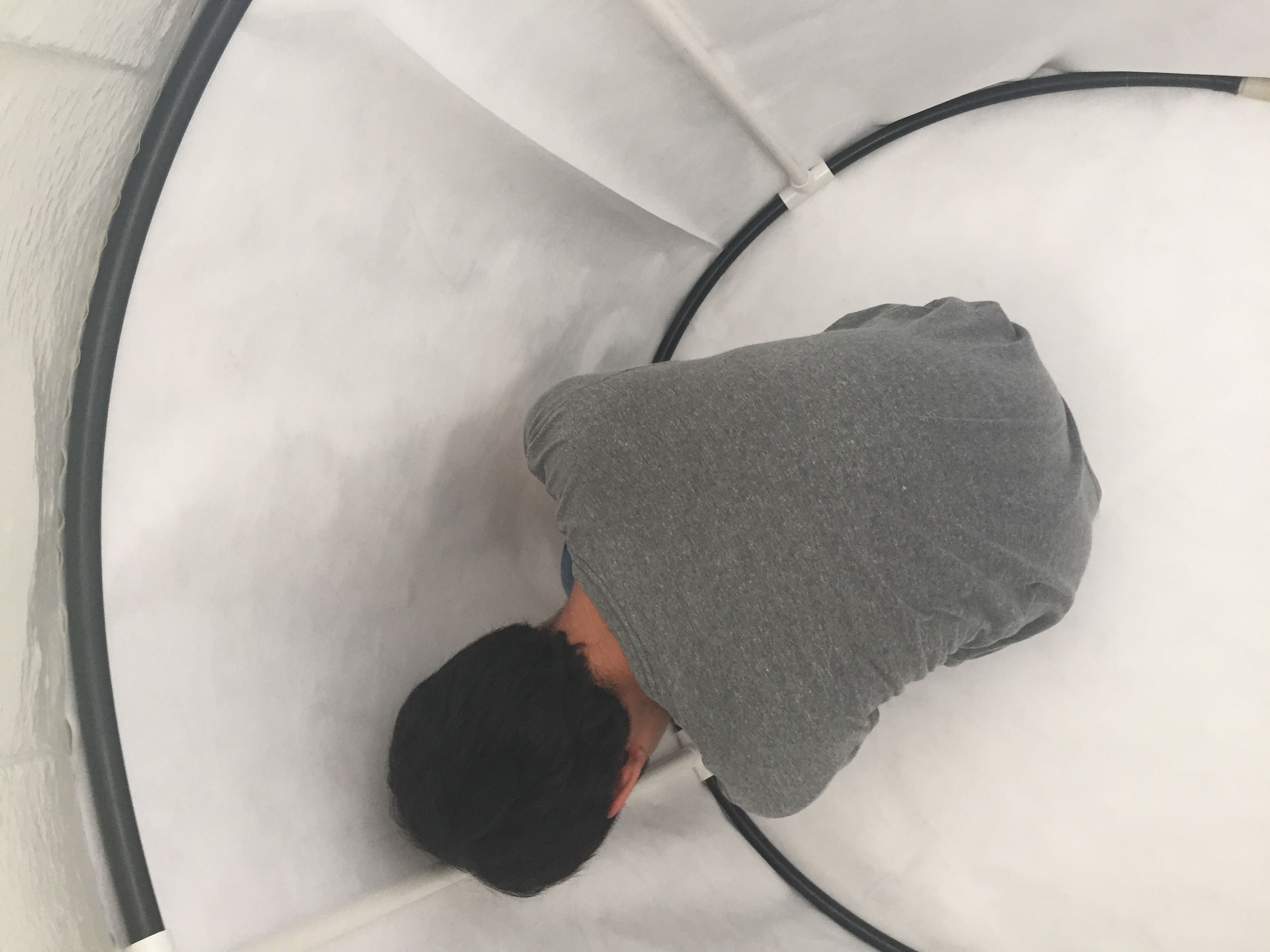}%
\figcaption{\emph{Inner surface of LAGO-Chiapas tank, being covered with Tyvek paper, using a pvc structure.}}}
\label{fig:6}
\hspace{2.6pc}
\centering
\myfigure{\includegraphics[scale=0.05]{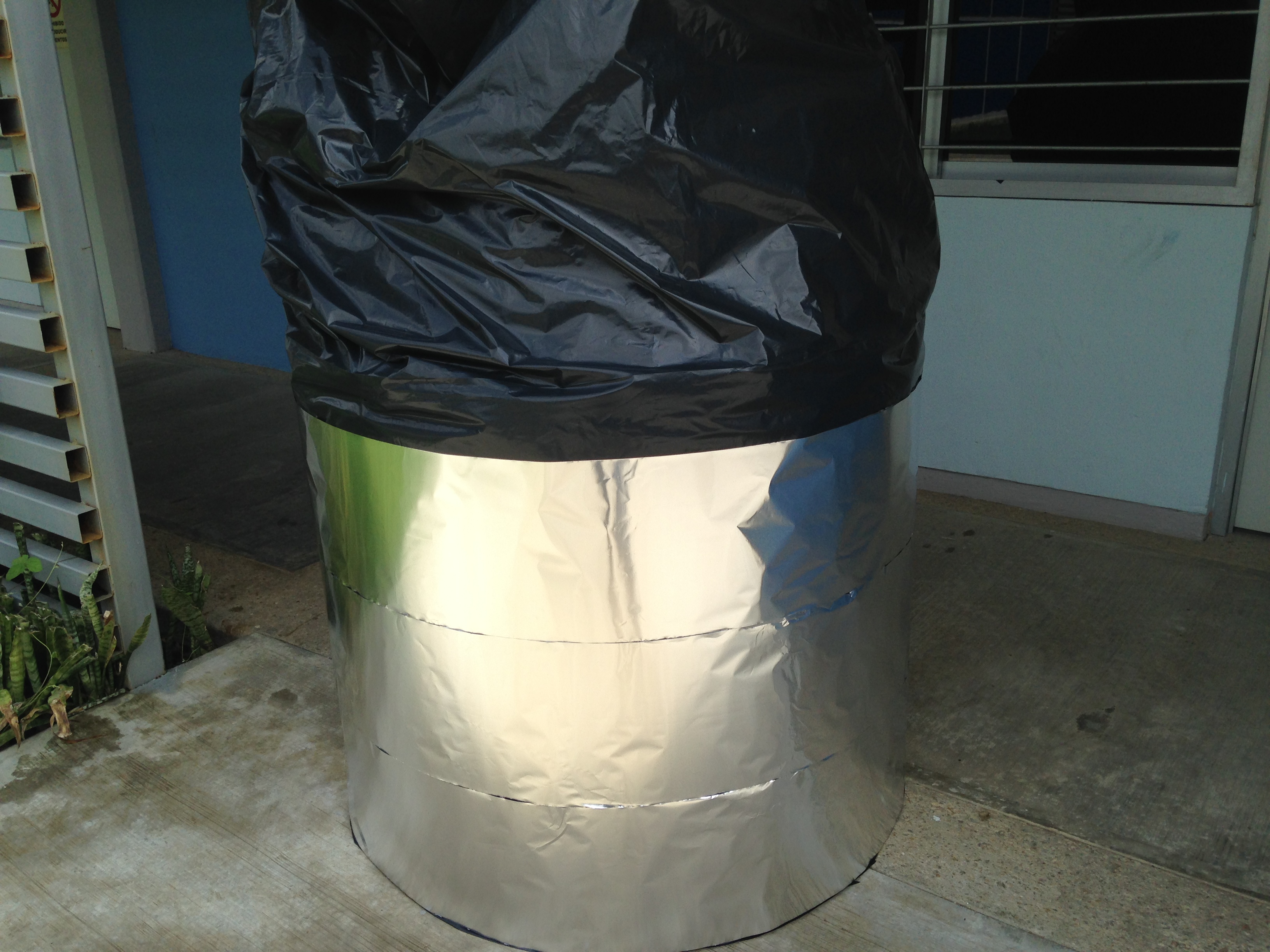}%
\figcaption{\emph{Outer surface of LAGO-Chiapas tank, the side wall is covered with ten plastic layers plus two of aluminium plus two more plastic layers at the end.}}}
\label{fig7}

\end{multicols}
\vspace{-1.2pc}
The acquisition is developed through a CAEN acquisition card (see \figurename{ 8}). CAEN provides several cards, some specialized in the acquisition of signals from PMT's such as the V1720 digitizer card, the V6533 high voltage source card, theV1718 master control card, among others. The programming of the architecture of the card must be done so that it works according to our needs, adapting the standard physical (types and forms of connectors, etc.), logical (signal levels, distribution of functions in different blocks, form of access to the data bus, interruptions, etc.)  and communication (bandwidth, multiplexed or non-multiplexed transfer types, number of bits in the transfer, etc.) aspects of the card. So far the master control board, the high voltage source cards and a digitizer have been programmed. The latter is the one that will perform the acquisition of signals. Programs in g++ using a library of  Application Programming Interface (API) functions provided by the manufacturer are developed. \figurename{ 9} shows the image of one of the PMT's (Photonis 9 "XP1805) that are in the FCFM. At the moment it is already possible to see signals from atmospheric muons out of the tank in the PMT (see \figurename{ 10} ). A RedPitaya \cite{Pitaya}  card is also available,  this is the card which is expected to be used in all LAGO sites in order to standardize the performance of the instruments.

\begin{multicols}{2}

\myfigure{\includegraphics[scale=1.3]{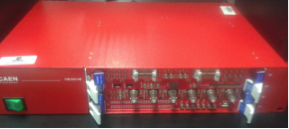}%
\figcaption{\emph{CAEN signal acquisition equipment}}}
\label{fig:8}
\hspace{1.5pc}
\centering
\myfigure{\includegraphics[scale=1.76]{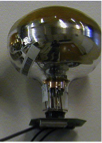}%
\figcaption{\emph{Photonis 9 "XP1805 PMT}}}
\label{fig9}
\vspace{-0.95pc}
\myfigure{\includegraphics[scale=0.14]{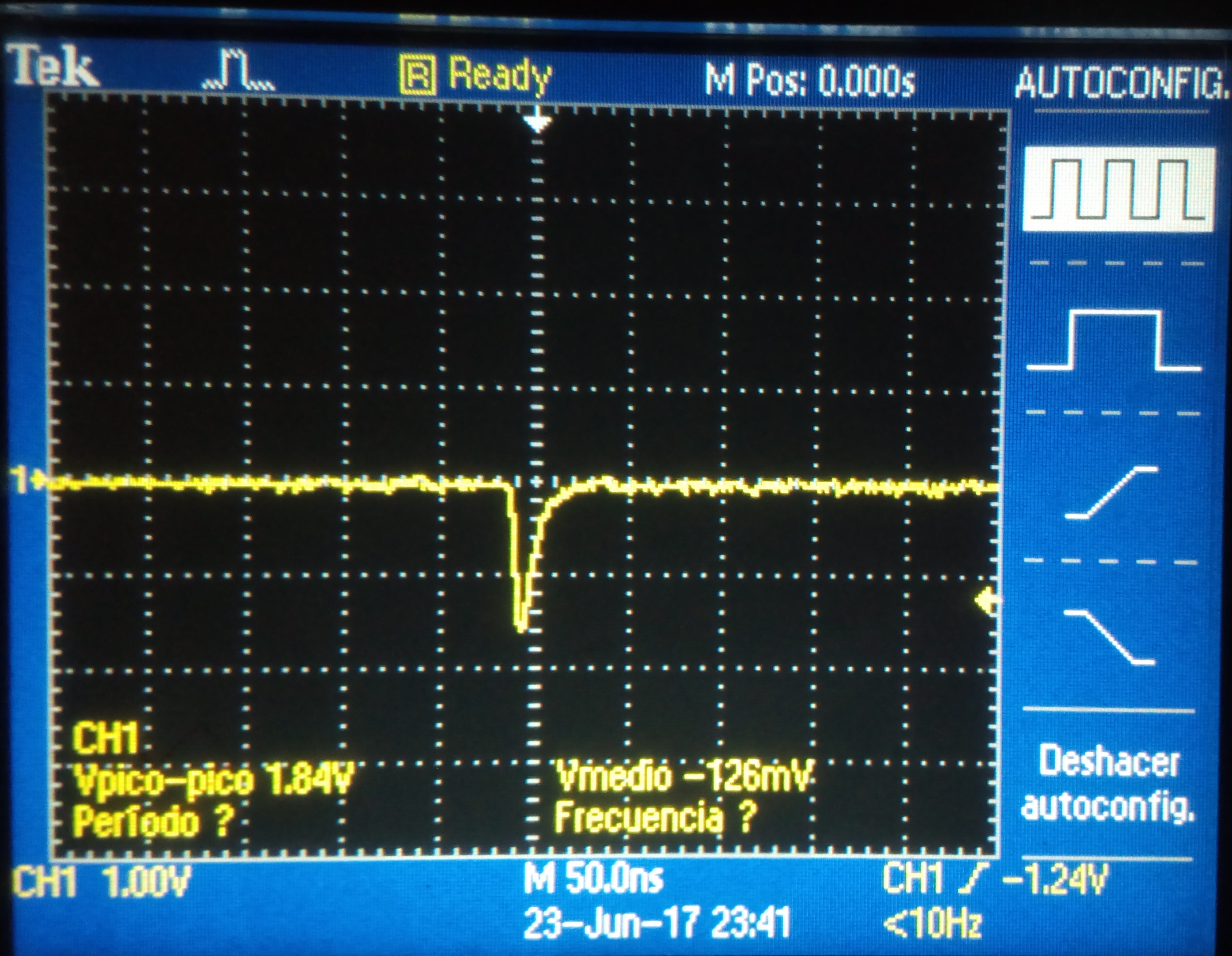}%
\figcaption{\emph{One Muonic signal seen by the PMT on the screen of an osciloscope. Y axis shows the output voltage (in $V$) and the X axis is the time in $ns$}}}
\label{fig10}

\end{multicols}

The next steps  to be taken are to perform tests of atmospheric muon rates to process the information obtained by the acquisition card and to compare it with the measurements from the Escaramujo device. Once the tests are done, the PMT is going to be installed in the tank with pure water and the calibration process, using Escaramujo, will be performed to finally obtain the first signals from the LAGO-Chiapas tank.

\section{Conclusions}
UNACH's involvement in the LAGO project is being crucial to learn on a first-hand basis, the development of a high-quality astroparticle detector. Students interested in this field of research are acquiring invaluable experience and obtaining the scientific and technological knowledge necessary to develop experiments of great impact at international level. This project will allow them to continue their preparation and establish collaborations with researchers from other prestigious institutions worldwide. The contribution in basic science that can be obtained from this effort will have transcendence at the international level in the field of astroparticles and great impact in the development of science in the region.
\section{Acknowledgments}
We gratefully acknowledge projects FECES 2015, Ciencia B\'asica CONACyT 243290 and UNACH-PTC-166-2017  for the financial support to develop the presented work.


\end{document}